\documentclass[12pt]{iopart}

\usepackage{iopams}
\usepackage[dvips]{graphicx}

\begin{document}

\title{Nonextensive effects in the Nambu--Jona-Lasinio model
of QCD}

\author{Jacek Ro\.zynek}
\address{The Andrzej So{\l}tan Institute for Nuclear Studies,
Ho\.{z}a 69, 00681, Warsaw, Poland} \ead{rozynek@fuw.edu.pl}
\author{Grzegorz Wilk}
\address{The Andrzej So{\l}tan Institute for Nuclear Studies,
Ho\.{z}a 69, 00681, Warsaw, Poland} \ead{wilk@fuw.edu.pl}

\begin{abstract}
We present a nonextensive version of the QCD-based Nambu -
Jona-Lasinio (NJL) model of a many-body field theory describing
the behavior of strongly interacting matter. It is based on the
nonextensive generalization of Boltzmann-Gibbs (BG) statistical
mechanics used in the NJL model, which was taken in the form
proposed by Tsallis characterized by a dimensionless
nonextensivity parameter $q$ (for $q \rightarrow 1$ one recovers
the usual BG case). This new phenomenological parameter accounts
summarily for all possible  effects resulting in a departure from
the conditions required by application of the BG approach, and
allows for a simple phenomenological check of the sensitivity of
the usual NJL model to such effects (in particular to fluctuations
of temperature and correlations in a system of quarks). As an
example, we discuss the sensitivity of such a $q$-NJL model to the
departures from the NJL form, both for  $ q> 1$ and $ q < 1$
cases, for such observables as the temperature dependencies of
chiral symmetry restoration, masses of $\pi$ and $\sigma$ mesons
and characteristic features of spinodal decomposition.
\end{abstract}


\maketitle

\section{\label{sec:I}Introduction}

In all standard studies of high energy collisions and properties
of nuclear matter one often uses a statistical approach based on
Boltzmann-Gibbs (BG) statistics. However, such an approach is,
strictly speaking, only correct when the corresponding heat bath
is homogeneous and infinite. These conditions are not met in
realistic situations in which one encounters some inherent
problems arising, for example, from the smallness of the collision
system and from its rapid evolution. These, among other things,
render the spatial configuration of the system being far from
uniform and prevent global equilibrium from to be established (cf.
\cite{departure} and references therein). As a result, some
quantities become non extensive and develop power-law tailed
rather than exponential distributions.

However, one can still use a reasonably simple statistical
approach provided it will be based on a nonextensive extension of
the usual BG statistics known as $q$-statistics (cf.
\cite{T,applications,EPJA,AL} and references therein). The new
phenomenological nonextensivity parameter $q$ occurring here
accounts for all possible dynamical factors violating assumptions
of the usual BG statistics which is recovered in the limit of $q
\rightarrow 1 $. Because it enters into the respective formulas of
the particular dynamical model used for a given investigation, it
allows for the phenomenological check of the stability of the
model considered against possible deviations from the BG
approach\footnote{Applications of the nonextensive approach are
numerous and cover all branches of physics \cite{applications}.
For those in high energy multiparticle production processes see
\cite{EPJA,qHydro,qH} whereas \cite{AL,qBiro,Drago} deal with
different aspects of nuclear and quark matter. The nonextensive
framework can also be derived from a special treatment of kinetic
theory investigating complex systems in their nonequilibrium
stationary states \cite{qBiro,stationary,Kaniadakis}. In
\cite{MaxEnt_K} one can find updated discussion of power-law
tailed distributions emerging from an maximum entropy principle.
Some examples of more specialized topics can be found in
\cite{todos} and references therein.}.

In what concerns possible physical interpretations of the
parameter $q$, the most popular one is that for the $q > 1$ case,
$q - 1$ is a measure of intrinsic fluctuations of the temperature
in the system considered \cite{WW,BJ,EPJA} whereas  $q < 1$ is
usually attributed to some specific correlations limiting the
available phase space \cite{Kodama} or to the possible fractality
of the allowed phase space \cite{fractal} (still other possible
interpretations were considered in \cite{todos}).  As a result of
application of the $q$-statistics, one gets a characteristic
power-law distribution in energy-momentum \cite{EPJA} and specific
$q$-versions of the Fermi-Dirac (FD) distribution \cite{TPM,TPM1}
(see also \cite{DBG,Drago}).

Notice that in $q$-statistics we do not specify what is the
dynamical origin of these intrinsic fluctuations or specific
correlations. It is expected that every piece of a new dynamical
knowledge accumulated during systematic studies of the respective
processes substantially lowers the values of the parameter $|q-1|$
needed to fit experimental data. This was confirmed in \cite{qH}
when investigating transverse momenta distributions in heavy ion
collisions, namely the gradual accounting for the intrinsic
dynamical fluctuations in the hadronizing system by switching from
pure statistical approach to the modified Hagedorn formula
including temperature fluctuations \cite{qH}, resulted in sizeable
decreasing of $q-1$. This means, therefore, that when one reaches
in such a procedure the value $q = 1$, it should signal that all
dynamical effects spoiling the initially assumed BG approach have
already been successfully accounted for.

Recently, $q$-statistics has been applied to the Walecka many-body
field theory \cite{SW} (known as quantum hadrodynamics or QHD-I)
\cite{Pereira}. It resulted, among other things, in the
enhancement of the scalar and vector meson fields in nuclear
matter, in diminishing of the nucleon effective mass and in
hardening of the nuclear equation of state (only the $q > 1$ case
was considered there).

In this paper we shall present a nonextensive version of the
QCD-based Nambu - Jona-Lasinio (NJL) model of a many-body field
theory describing the behavior of strongly interacting matter by
accordingly modifying the NJL model recently presented in
\cite{Sousa}. This means than that, unlike in \cite{Pereira}, we
shall work on the quark rather than the hadronic level. Also,
unlike in \cite{Pereira}, we shall consider both the $q > 1$ and
$q < 1$ cases. This will allow us to discuss the $q$-dependence of
the chiral phase transition in dense quark matter, in particular
the quark condensates and the effective quark masses. Their
influence on the masses of $\pi$ and $\sigma$ mesons and on the
spinodal decomposition will be also presented. In the present work
we shall limit ourselves to investigation of the response of these
two elements to the departure from the usual BG approach and
confront the obtained results with a possible dynamical
explanations \footnote{Actually, in \cite{Sousa} the systematic
investigation of the phase diagram of strongly interacting matter
as a function of temperature and chemical potential was
undertaken, which we shall not repeat here in the nonextensive
version as this would take us outside the limited scope of the
present paper.  We plan to consider it elsewhere.}.

Our paper is organized as follows: In Section \ref{sec:II} we
provide a short reminder of the basic features of the NJL model
used in \cite{Sousa}. Section \ref{sec:III} contains a formulation
of the nonextensive version of the NJL model (the $q$-NJL) whereas
our results are presented in Section \ref{sec:Results} where we
discuss the influence of nonextensive statistics on chiral
symmetry restoration in $q$-NJL (Section \ref{sec:Chiral}) and the
$q$ version of the spinodal decomposition (Section
\ref{sec:Spinodal}). We close with a summary and conclusions in
Section \ref{sec:IV}.

\section{\label{sec:II}Basic elements of the NJL model}

The SU$(3)$ NJL model with U$(1)_A$ symmetry breaking was first
formulated and discussed in \cite{Bernard1} and the first
bosonized version of the NJL that obeys all strictures of chiral
symmetry was discussed in \cite{Bernard2}. We start with
recollecting some basic formulas concerning the NJL model used in
\cite{Sousa} (see also \cite{NJL,HK,Kle,Bub}). They used the usual
lagrangian of the NJL model, invariant (except of the current
quarks mass term) under the chiral SU$_L(3)\otimes$SU$_R(3)$
transformations (described by coupling constant $g_S$) and
containing a term breaking the U$_A(1)$ symmetry, which reflects
the axial anomaly in QCD (described by coupling constant $g_D$).
When put in a form suitable for the bosonization procedure (with
four quark interaction only) it results in the following effective
lagrangian:
\begin{eqnarray}
{\cal L}_{eff} = \bar{q}\left(i {\gamma}^{\mu}\partial_\mu
-\hat{m}\right)q\! +\! S_{ab}\left[\left(\bar{q}\lambda^a q
\right)\left( \bar{q}\lambda^b q \right)\right]\! + \!
P_{ab}\left[\left(\bar{q} i\gamma_5 \lambda^a q \right)\left(
\bar{q} i\gamma_5\lambda^b q \right)\right], \label{lagr_eff}
\end{eqnarray}
where $\hat{m} = {\rm diag}\left(m_u, m_d, m_s\right)$ and
$S_{ab}$ and $P_{ab}$ are projectors,
\begin{eqnarray}
S_{ab} &=& g_S \delta_{ab} + g_D D_{abc}\left\langle \bar{q}
\lambda^c
q\right\rangle, \label{sab}\\
P_{ab} &=& g_S \delta_{ab} - g_D D_{abc}\left\langle \bar{q}
\lambda^c q\right\rangle \label{pab}
\end{eqnarray}
with $D_{abc}$ being the SU(3) structure constants $d_{abc}$ for
$a,b,c=(1,2,\dots,8)$ whereas $D_{0ab} = -\delta_{ab}/\sqrt{6}$
and $D_{000} = \sqrt{2/3}$. We work with $q = (u,d,s)$ quark
fields with three flavors, $N_f = 3$, and three colors, $N_c = 3$,
$\lambda^a$ are the Gell-Mann matrices, $a = 0,1,\ldots , 8$ and
${\lambda^0=\sqrt{\frac{2}{3}} \, {\bf I}}$. Integrating over the
quark fields in the functional integral with ${\cal L}_{eff}$ one
gets an effective action expressed by the natural degrees of
freedom of low energy QCD in the mesonic sector, namely $\sigma$
and $\varphi $ (the notation $\mbox{Tr}$ stands for taking trace
over indices $N_f$ and $N_c$ and integrating over momentum) :
\begin{eqnarray}
W_{eff}[\varphi,\sigma] &=& -\frac{1}{2}\left(
\sigma^{a}S_{ab}^{-1} \sigma^{b}\right)  - \frac{1}{2}\left(
\varphi^{a}P_{ab}^{-1}\varphi
^{b}\right) - \label{action}\\
&-& i\mbox{Tr}\,\mbox{ln}\Bigl[i\gamma^{\mu}\partial_{\mu}-\hat{m}%
+\sigma_{a}\lambda^{a}+(i\gamma_{5})(\varphi_{a}\lambda^{a})\Bigr]\,.
\nonumber
\end{eqnarray}

The first variation of $W_{eff}$ leads to the gap equations for
the constituent quark masses $M_i$:
\begin{eqnarray}
 M_i = m_i - 2g_{_S} \big <\bar{q_i}q_i \big > -2g_{_D}\big
 <\bar{q_j}q_j\big > \big <\bar{q_k}q_k \big >\,,\label{gap}
 \end{eqnarray}
with cyclic permutation of $i,j,k =u,d,s$ and with the quark
condensates given by $\big <\bar{q}_i q_i \big > = -i \mbox{Tr}[
S_i(p)]$ ($S_i(p)$ is the quark Green function); $m_i$ denotes the
current mass of quark of flavor $i$ (notice that nonzero $g_D$
introduces mixing between different flavors).

Let us consider a system of volume $V$, temperature $T$ and the
$i^{th}$ quark chemical potential $\mu_i$ characterized by the
baryonic thermodynamic potential of the grand canonical ensemble
(with quark density equal to $\rho_i = N_i/V$, the baryonic
chemical potential $\mu_B= \frac{1}{3} (\mu_u+\mu_d+\mu_s)$ and
the baryonic matter density as $\rho_B =
\frac{1}{3}(\rho_u+\rho_d+\rho_s)$),
\begin{equation}
\Omega (T, V, \mu_i )= E- TS - \sum_{i=u,d,s} \mu _{i} N_{i} .
\label{tpot}
\end{equation}
The internal energy, $E$, the entropy, $S$, and the particle
number, $N_i$, are given by \cite{Sousa,Costa} (here $E_i =
\sqrt{M_i^2 + p^2}$):
\begin{eqnarray}
E &=&- \frac{ N_c}{\pi^2} V\sum_{i=u,d,s}\left[
   \int p^2 dp  \frac{p^2 + m_{i} M_{i}}{E_{i}}
   (1 - n_{i}- \bar{n}_{i}) \right] - \nonumber\\
   && - g_{S} V \sum_{i=u,d,s}\, \left(\big <
\bar{q}_{i}q_{i}\big > \right)^{2}
   - 2 g_{D}V \big < \bar{u}u\big > \big < \bar{d}d\big > \big <
\bar{s}s\big > , \label{energy} \\
 S &=& \! -\frac{ N_c}{\pi^2} V \sum_{i=u,d,s}\int p^2 dp \cdot
 \tilde{S}, \label{entropy}\\
 && {\rm where}\quad \tilde{S} =  \bigl[ n_{i} \ln n_{i}+(1-n_{i})\ln (1-n_{i})
   \bigr]\!\! +\!\! \bigl[ n_{i}\rightarrow 1 - \bar n_{i} \bigr],\nonumber\\
N_i &=& \frac{ N_c}{\pi^2} V \int p^2 dp
  \left( n_{i}-\bar n_{i} \right) \label{number}.
\end{eqnarray}
The quark and antiquark occupation numbers, $n_i$ and $\bar n_i$,
are
\begin{equation}
n_{i}= \frac{1}{\exp\left[\beta \left(E_{i} -
\mu_{i}\right)\right] + 1},\qquad \qquad \bar n_{i} =
\frac{1}{\exp\left[ \left( \beta(E_{i} + \mu_{i} \right)\right] +
1}.
\end{equation}
With these occupation numbers one can now calculate values of the
quark condensates present in Eq. (\ref{gap}),
\begin{eqnarray}
\big <\bar{q}_i  q_i \big> \!= \!\ - \frac{ N_c}{\pi^2} \!\!\!
\sum_{i=u,d,s}\left[ \int \frac{p^2M_i}{E_i} (1\,-\,n_{i}-\bar
n_{i})\right]dp .\label{gap1}
\end{eqnarray}
Eqs. (\ref{gap}) and (\ref{gap1}) form a self consistent set of
equations from which one gets the effective quark masses $M_i$ and
values of the corresponding quark condensates (once a temperature
and chemical potential are given).

The values of the pressure, $P$, and the energy density,
$\epsilon$,
\begin{equation} \label{p}
   P(\mu_i, T) = - \frac{\Omega(\mu_i, T)}{V},\qquad \qquad
   \epsilon(\mu_i, T) = \frac{E(\mu_i, T)}{V}
\end{equation}
are defined such that $P(0,0) = \epsilon(0,0) = 0$.

In order to illustrate the $q$-dependence of the chiral phase
transition for zero chemical potential, in the present work we are
only concerned with $\pi^0$ and $\sigma$ mesons. Their effective
masses can be obtained from the effective action (\ref{action}) by
expanding it over meson fields and calculating the respective
propagators. In the case of a $\sigma$ meson one must also account
for its matrix structure in isospin space cf., \cite{Sousa}. And
thus the mass of the $\pi^0$ meson is be determined by the
condition:
\begin{equation}
D_{\pi^0}^{-1}(M_{\pi^0},\mathbf{0})=0.\label{mesq}
\end{equation}
where $D_{\pi^0}^{-1}$ is the inverse of the meson $\pi^0$
propagator,
\begin{eqnarray}
D^{-1}_{\pi^0} (P) &=& 1-P_{\pi^0} J_{uu}^P (P), \\
&& P_{\pi^0}=g_{S}+g_{D}\left\langle\bar{q}_{s}q_{s}\right\rangle
\end{eqnarray}
and operator $J_{uu}$ is given by well defined integrals \cite{HK}
(cf., also appendix of ref.(\cite{Sousa})). The procedure for
obtaining the effective mass of the $\sigma$ meson is analogous
with the only difference that solving the condition
$D_{\sigma}^{-1}(M_{\sigma},\mathbf{0}) = 0 $ we use approximate
form of $\sigma$ propagator (see \cite{HK} for details). Finally,
the model is fixed by the coupling constants, $g_S$ and $ g_D$,
the current quark masses, $m_i$, and the cutoff $\Lambda$, which
is used to regularize the momentum space integrals \footnote{For
numerical calculations we use the same parameter set as that in
\cite{Sousa}: $m_u = m_d = 5.5$ MeV, $m_s = 140.7$ MeV, $g_S
\Lambda^2 =   3.67$, $g_D \Lambda^5 = - 12.36$ and $\Lambda =
602.3$ MeV. It has been determined by fixing the values $M_{\pi} =
135.0$ MeV, $M_K   = 497.7$ MeV, $f_\pi =  92.4$ MeV, and
$M_{\eta'}= 960.8$ MeV. For the quark condensates at $T=0$ we
obtain: $\left\langle \bar{q}_{u}\,q_u\right\rangle =
\left\langle\bar{q}_{d}q_d\right\rangle = - (241.9 \mbox{ MeV})^3$
and $\left\langle\bar{q}_{s}q_s\right\rangle = - (257.7 \mbox{
MeV})^3$, and for the constituent quark masses $M_u = M_d = 367.7$
MeV and $M_s= 549.5$ MeV.}.

\section{\label{sec:III}Nonextensive NJL model - $q$-NJL}

\subsection{\label{sec:Why}Motivation}

As a motivation for study of nonextensive version of the NJL
model, the $q$-NJL model, let us notice the following. The NJL
model \cite{Sousa} is formulated in the grand canonical ensemble
and assumes the additivity  of some thermodynamical properties,
especially entropy. This is a very strong approximation for the
system under the phase transition where long range correlations or
fluctuations are very important. One could, alternatively,
consider equilibrium statistics using microcanonical ensembles of
Hamiltonian systems, whereas canonical ensembles fail in the most
interesting, mostly inhomogeneous, situations like phase
separations or away from the thermodynamic limit \cite{Gross}. The
alternative way to describe the non additivity of interacting
systems which have long range correlations (including long range
microscopic memory) or long range microscopic interactions is to
use $q$-statistics \cite{Tq}.

Let us illustrate this with two examples. At first notice that in
the NJL model (which in the mean field approximation is given by
the effective lagrangian (\ref{lagr_eff})) one introduces a strong
attractive interaction between a quark and antiquark represented
by couplings $g_S$ and $g_D$; usually assumed to be independent of
the temperature $T$. However, this interaction induces the
instability of the Fock vacuum of the massless quarks which, in
turn, results in the non-perturbative ground state with nonzero
$(q\bar{q})$ condensates and in the breaking of chiral symmetry
endowing constituent quarks with finite masses. This effect takes
place in some range of temperatures so as to control it one allows
in some cases for a temperature dependent coupling constant $g_D$
as, for example, \cite{HK}. It was assumed there that $g_D$,
corresponding to breaking of axial symmetry $U_A(1)$, is given by
\begin{equation}
g_D(T) = g_D(T=0) \exp\left[ - \left(
\frac{T}{T_0}\right)^2\right], \label{memory}
\end{equation}
where $T_0$ is a parameter ($T_0 = 100$ MeV in \cite{HK}). As
shown in \cite{HK}, depending on the assumed value of $T_0$,
chiral symmetry starts earlier. The retardation effect introduced
by Eq. (\ref{memory}) violates the simple extensivity of the
system, therefore it calls for an effective nonextensive
description provided by $q$-statistics \cite{T}.

The second example concerns the description of the spinodal region
in the NJL models in which one observes a coexistence of two
phases: $(a)$ - the phase with broken chiral symmetry and with
massive quarks ($m \sim 300$ MeV) and large negative $q\bar{q}$
condensates which constitute the physical vacuum, it develops for
small density; $(b)$ - for high density the $q\bar{q}$ condensates
disappear and quarks are almost massless ($m \sim 5$ MeV). The
highest point on the temperature scale of the coexistence curve,
$T_{crit}$, is the critical point. Of special importance is the
fact that, within $q$-statistic, one can discuss the occurrence of
negative specific heat in a nonextensive system which has an
equilibrium second order phase transition \cite{NegC}. According
to this analysis, the specific heat is negative in a transient
regime and corresponds to meta-stable states. Exactly such
metastable states are observed in the NJL model during the
spinodal phase transition below the critical temperature
\cite{Sousa}.

Finally, let us notice that the NJL model does not contain color
and therefore does not produce confinement. Therefore, resigning
from the assumption of additivity in this case and introducing a
description based on the nonextensive approach, which, according
to \cite{qBiro}, can be understood as containing some residual
interactions between considered objects (here quarks) seems to be
an interesting and promising possibility.

\subsection{\label{sec:Formulation}Formulation of the $q$-NJL}

The nonextensive statistical mechanics proposed by Tsallis
\cite{T} generalizes the usual BG statistical mechanics in that
entropy function (we use convention that Boltzmann constant is set
equal to unity),
\begin{equation}
S_{BG} = - \sum_{i=1}^{W}p_i\ln p_i  \Longrightarrow  S_{q} = -
\sum_{i=1}^W p_i^q \ln_q p_i,
\end{equation}
$S_q \rightarrow S_{q=1} = S_{BG}$ for $q \rightarrow 1$. Here,
$q$ is the nonextensive parameter and $\ln_q p = \left[ p^{1-q}-1
\right]/(1 - q)$. The additivity for two independent subsystems A
and B (i.e., such that $p^{A\oplus B} = p^{A}\cdot p^{B}$) is now
lost and takes the form:
\begin{equation}
{S^{A\oplus B}_{q}}= {S_{q}^{A}} + {S^{B}_{q}}+(1-q){S_{q}^{A}}
{S^{B}_{q}}, \label{eq:S-q}
\end{equation}
they are called nonextensive \footnote{It is worth knowing that
for subsystems with some special probability correlations, it is
the BG entropy for which extensivity is not valid and is restored
only for $q \neq 1$ (one refers to such systems as nonextensive
\cite{tsallis05}).}.

The relevant point for further consideration is the $q$-form of
quantum distributions for fermions $(+1)$ and bosons $(-1)$,
which, following \cite{TPM} (and \cite{Pereira}) we shall take as:
\begin{eqnarray}
n_{qi} &=& \frac{1}{\tilde{e}_q(\beta(E_i - \mu_i))\pm
1},\label{nq} \label{TPM}
\end{eqnarray}
where, for $q > 1$ considered there,
\begin{eqnarray}
\tilde{e}_q(x) &=& \left\{
\begin{array}{l}
~[1+(q-1)x]^{\frac{1}{q-1}}\quad {\rm if}\quad x > 0 \\
\\
~[1+(1-q)x]^{\frac{1}{1-q}}\quad {\rm if}\quad x\leq 0 \\
\end{array}
\right. \label{q>1}
\end{eqnarray}
and $x = \beta(E -\mu)$. The $q < 1$ case was not considered in
\cite{Pereira} whereas \cite{TPM} advocated use of the usual
Tsallis cut prescription in this case, i.e., to allow for a given
$q < 1$ only for such values of $(E, \mu, \beta$) for which $[1 +
(1-q)x] \ge 0$. However, in this case we found it more suitable to
adopt in this case mirror reflection of Eq. (\ref{q>1}), i.e.,
that for $ q < 1$ one has:
\begin{eqnarray}
\tilde{e}_q(x) &=& \left\{
\begin{array}{l}
~[1+(q-1)x]^{\frac{1}{q-1}}\quad {\rm if}\quad x \leq 0 \\
\\
~[1+(1-q)x]^{\frac{1}{1-q}}\quad {\rm if}\quad x > 0 \\
\end{array}
\right. .\label{q<1}
\end{eqnarray}
This is because only then can one treat consistently on the same
footing (and for all values of $x$) quarks and antiquarks, which
should show the particle-hole symmetry observed in the $q$-Fermi
distribution in plasma containing both particles and
antiparticles, namely that
\begin{equation}
n_q(E,\beta,\mu,q) = 1 - n_{2-q}(- E,\beta,-\mu).
\label{pap_symmetry}
\end{equation}
This means, therefore, that in a system containing both particles
and antiparticles (as in our case) both $q$ and $2 - q$ occur (or,
when expressed by a single $q$ only, that one can encounter both
$q > 1$ and $q <1$ at the same time). These dual possibilities
warn us that not only the $q > 1$ but also $ q < 1$ (or $(2 - q) >
1$ have physical meaning in the systems we are considering. This
differs our $q$-NJL model from the $q$-version of the QHD-I model
of \cite{Pereira}.

Notice that for $q\rightarrow 1$ one recovers the standard FD
distribution, $n(\mu,T)$. Actually, it is important to realize
that for $T\rightarrow0$ one always gets $n_q(\mu,T)\rightarrow
n(\mu,T)$, irrespectively of the value of $q$ \cite{Pereira}. This
means that we can expect any nonextensive signature only for high
enough temperatures.

In formulating the $q$-NJL, in what concerns calculations of the
modified FD distributions of quarks and antiquarks, we follow
essentially the steps undertaken in \cite{Pereira} where the
$q$-version of the Walecka model of nuclear matter has been
formulated and investigated using the cutoff prescription proposed
in \cite{TPM}. This allows a comparison of the $q$-version of both
approaches. Our $q$-NJL model is then obtained by replacing the
formulas of Section \ref{sec:II} with their $q$-counterparts in
what concerns the form of the FD distributions.  Additionally,
when calculating energies and condensates we follow
\cite{Drago,AL} and use the $q$-versions of energies and quark
condensates replacing Eqs. (\ref{energy}) and (\ref{gap1}) by:
\begin{eqnarray}
E_q\! &=&\! - \frac{ N_c}{\pi^2} V\!\!\!\sum_{i=u,d,s}\!\!\left[
   \int p^2 dp  \frac{p^2 + m_{i} M_{i}}{E_{i}}
   (1 - n^q_{qi}- \bar{n}^q_{qi}) \right] - \nonumber\\
   &&\!\! - g_{S} V\!\!\!\sum_{i=u,d,s}\!\! \left(\big <
\bar{q}_{i}q_{i}\big >_q \right)^{2}
   - 2 g_{D}V \big < \bar{u}u\big >_q \big < \bar{d}d\big >_q \big <
\bar{s}s\big >_q , \label{q_energy}
\end{eqnarray}
and
\begin{equation}
 \big <\bar{q}_i
q_i \big>_q \! = \!\ - \frac{ N_c}{\pi^2} \!\!\!
\sum_{i=u,d,s}\left[ \int \frac{p^2M_i}{E_i} (1\,-\,n^q_{qi}-
\bar{n}^q_{qi})\right]dp .\label{q_gap1}
\end{equation}
On the other hand, again following \cite{Drago,AL}, densities
which are given by the the $q$-version of Eq. (\ref{number}) are
calculated with $n_q$'s (not with $n_q^q$, as in (\ref{q_energy})
and in (\ref{q_gap1})). The pressure for given $q$ is calculated
using the above $E_q$ and the $q$-entropy version of Eq.
(\ref{entropy}) with (cf. \cite{TPM})
\begin{equation}
\tilde{S}_q\!\! =\!\! \left[ n^q_{qi} \ln_q n_{qi} +
(1-n_{qi})^q\ln_q (1-n_{qi}) \right] + \left\{ n_{qi}\rightarrow
1\! -\! \bar n_{qi} \right\}. \label{q entropy}
\end{equation}

\section{\label{sec:Results}Results}

In our study we concentrate on two features of the $q$-NJL model,
namely chiral symmetry restoration and spinodal decomposition, the
results for which we discus in what follows. Because our goal was
to demonstrate the sensitivity to the nonextensive effects
represented by the $|q - 1| \neq 0$, we do not reproduce here the
whole wealth of results provided in \cite{Sousa}, but concentrate
on the most representative.

\subsection{\label{sec:Chiral}Chiral symmetry restoration in the
$q$-NJL}

\begin{figure}[h]
\begin{center}
\includegraphics[width=7.7cm]{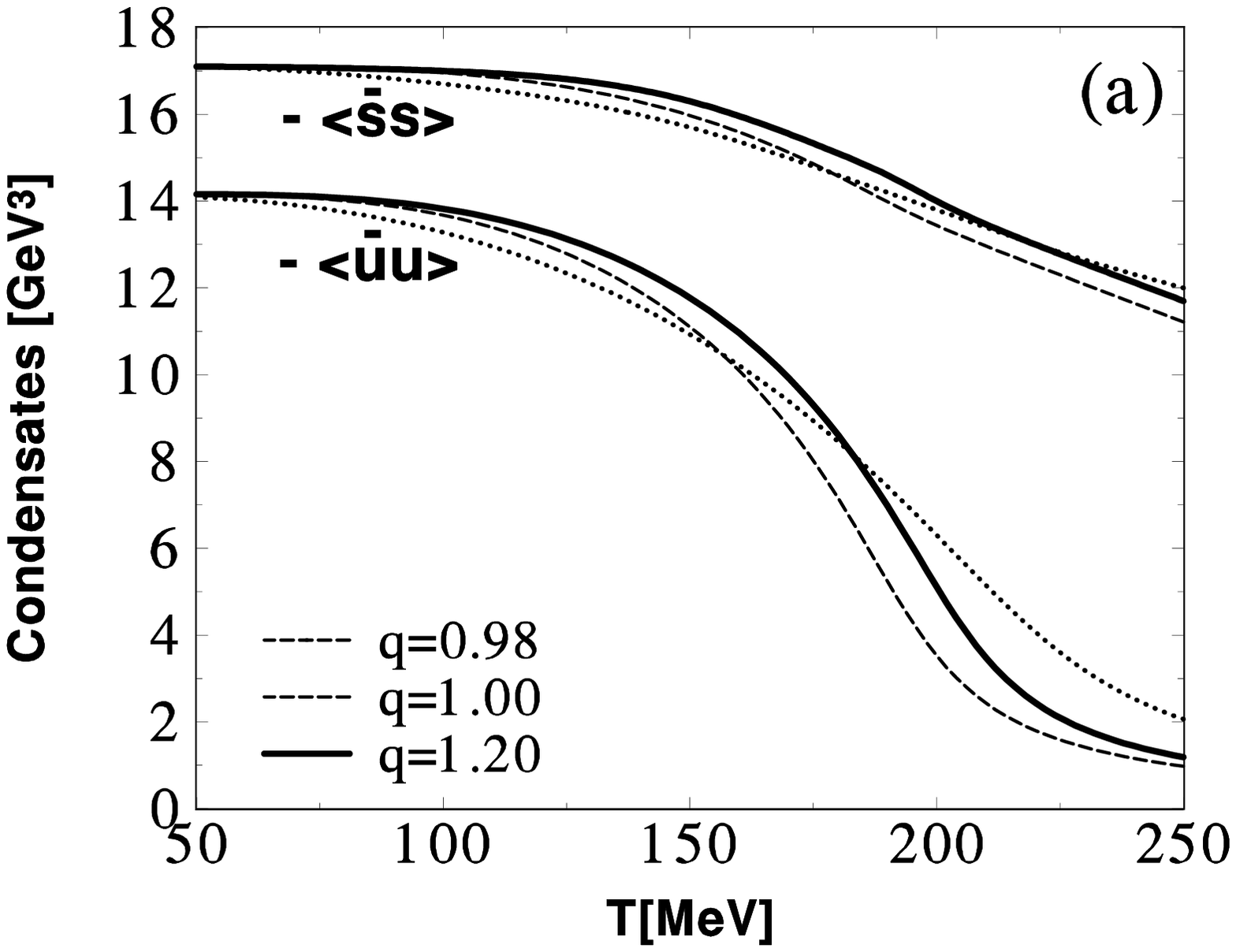}
\includegraphics[width=7.7cm]{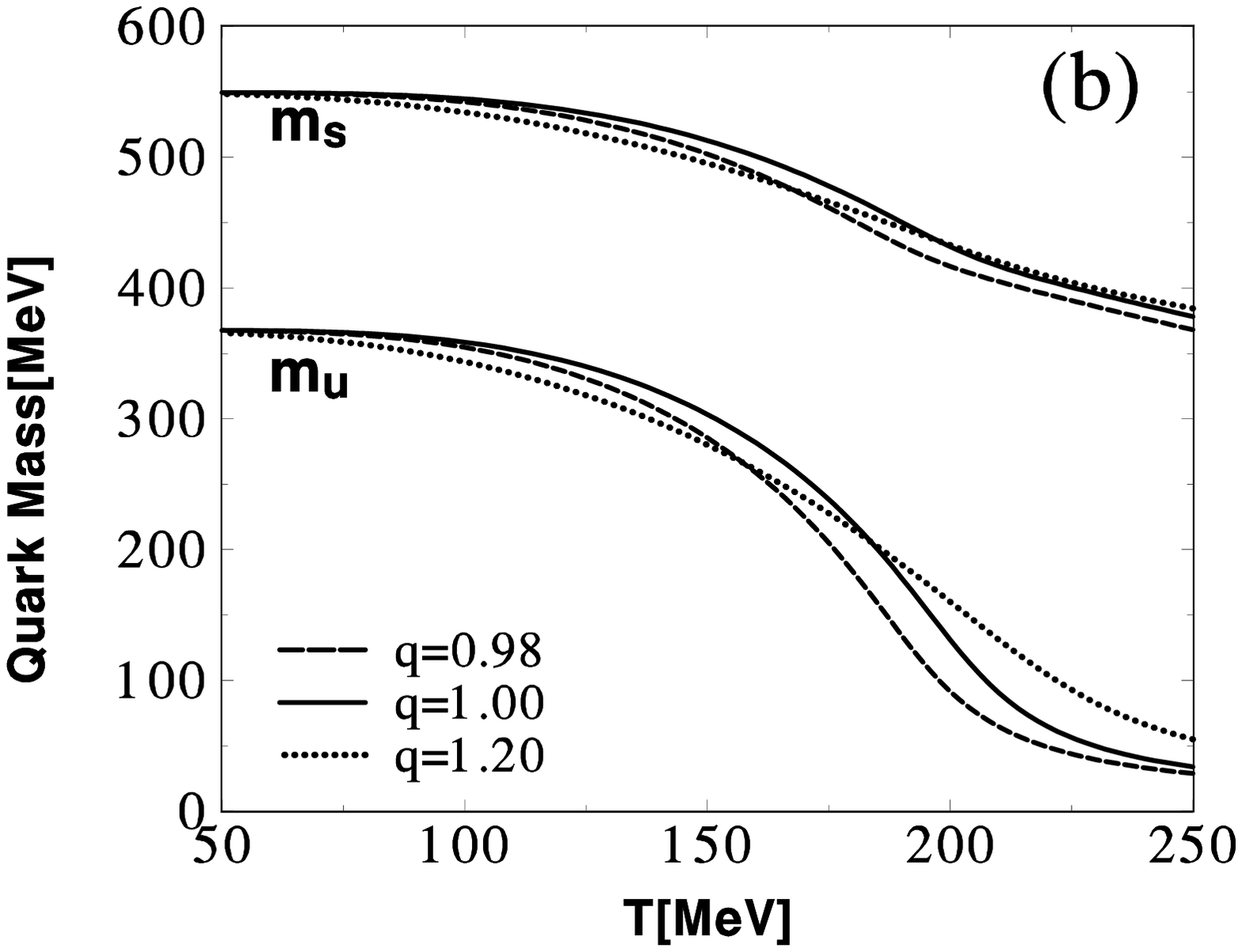}
\end{center}
\vspace{-0.5cm} \caption{(a) - quark condensates and (b) -
effective quark masses as functions of the temperature for
different values of the nonextensive parameter q (q=1 correspond
to Boltzmann Gibbs statistics).} \label{Fig:1}
\end{figure}

\begin{figure}[h]
\begin{center}
\includegraphics[width=8.5cm]{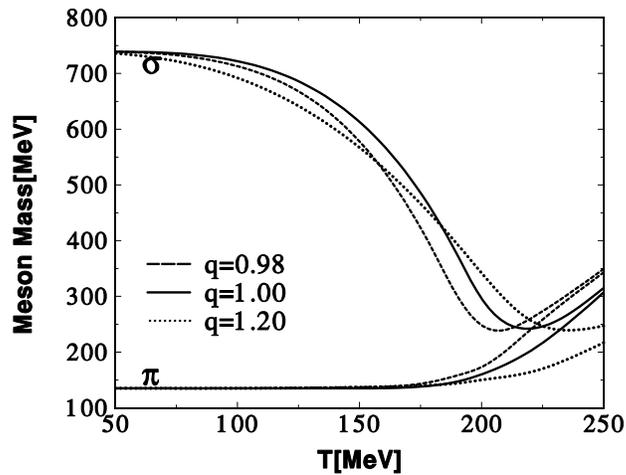}
\end{center}
\vspace{-0.8cm} \caption{Masses of $\pi$ and $ \sigma $ mesons as
functions of the temperature for different values of the
nonextensive parameter q (q=1 correspond to BG statistics).}
\label{Fig:2}
\end{figure}

Chiral symmetry restoration is best illustrated by the temperature
dependence of the quark condensates and effective masses of quarks
presented in Fig. \ref{Fig:1} for different values of parameter
$q$.
In addition, in Fig. \ref{Fig:2} the masses of $\pi$ and $\sigma$
mesons for different $q$ are presented as function of temperature
$T$ \footnote{There is still an ongoing discussion on the meaning
of the temperature in nonextensive systems. However, in our case
the small values of the parameter $q$ deduced from data allow us
to argue that, to first approximation, $T_q = T$ used here and in
\cite{Pereira}. In high energy physics it is just the hadronizing
temperature (and instead of the state of equilibrium one deals
there with some kind of stationary state). For a thorough
discussion of the temperature of nonextensive systems, see
\cite{Abe}.}. They were calculated assuming zero chemical
potentials and solving numerically the $q$-version of gap
equations (\ref{gap}) and (\ref{gap1}). There is a noticeable
difference for $q <1$ and $q > 1$ cases: whereas $q < 1$ leads to
chiral symmetry restoration starting earlier but in general
following the usual shape, for $q > 1$ it is smeared, starting
earlier and ending later. The effects caused by nonextensivity are
practically invisible for heavier quarks.
Also, as seen in Fig. 2, the degeneracies of $\pi$ and $\sigma$
mesons starts earlier for $q <1$ and much later for $q > 1$. Here
we derive the mass spectra with the help of the $q$-version of Eq.
(\ref{mesq}) solved numerically.  The temperature for which the
$\sigma$ mass reaches a minimum is shifted to smaller values for
$q <1$ and to larger ones for $q > 1$ by amount depending on the
value of $q - 1$. Also the final value of the masses is larger for
$q < 1$ and smaller for $q > 1$, the actual amount depends on the
value of $|q - 1|$. When interpreting $q > 1$ as a measure of
temperature fluctuations \cite{EPJA,WW,BJ}, this would mean that
fluctuations dilute the region where the chiral phase transition
takes place; it is especially visible for the $\sigma$ meson which
is still not saturated at the temperature $T = 50$ MeV, see Fig.
\ref{Fig:2}. On the other hand, correlations, which according to
\cite{Kodama,fractal} result in $ q < 1$, only shift the
condensates, quark masses and meson masses towards smaller
temperatures. In our case the supposed fluctuations and
correlations refer to quarks, not hadrons. It is interesting
therefore to note that in \cite{Pereira}, where degrees of freedom
used are nucleons, a similar shift towards smaller temperatures
occurs for $ q > 1$, i.e., for temperature fluctuations of
nucleons ($ q < 1$ is not considered there). Notice that, as
mentioned in Sec. \ref{sec:I}, we are not specifying here what are
the actual dynamical mechanisms behind such
fluctuations/correlations, we just model them by the parameter
$q$. An example of such dynamical effect, the temperature
dependence of the the respective coupling constants, is mentioned
above, cf. Eq. (\ref{memory}). The same remark also applies to the
discussion that follows.

\subsection{\label{sec:Spinodal}Influence of nonextensivity on
the spinodal decomposition in the $q$-NJL}

The next point we shall address is the influence of $q$-statistics
on the spinodal phase transition discussed in \cite{Sousa}. To
this end we must proceed to finite density calculations. As in
\cite{Sousa}, we assume chemical equilibrium in the form of $\mu_u
= \mu_d = \mu_s = \mu$. This allows us to work with nonzero $N_i$
in Eq. (\ref{number}) and thus with nonzero baryon density defined
as $\rho = \frac{1}{3}\sum_i N_i/V$.
\begin{figure}[h]
\begin{center}
\includegraphics[width=9.3cm]{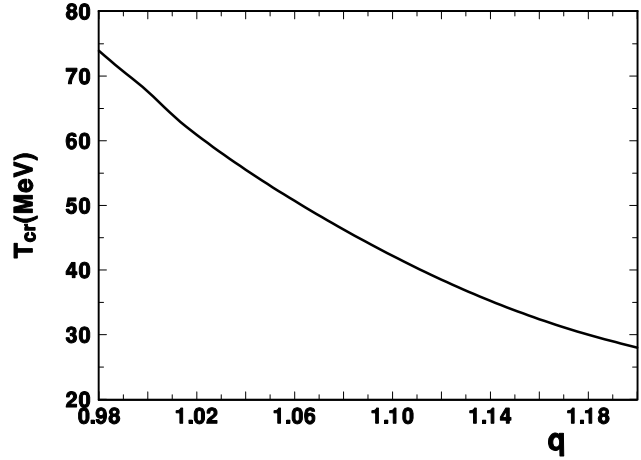}
\end{center}
\vspace{-1cm} \caption{Critical temperature $T_{cr}$ as function
of the nonextensivity parameter $q$ (in the range of $q$
considered here).} \label{Fig:3}
\end{figure}
\begin{figure}[h]
\begin{center}
\includegraphics[width=9cm]{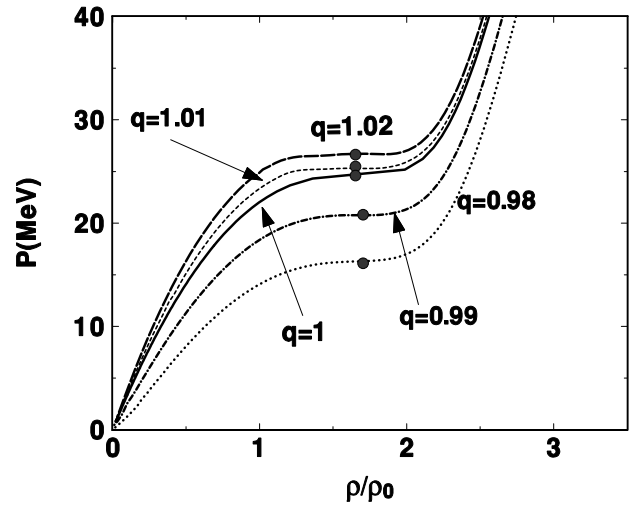}
\end{center}
\vspace{-0.5cm} \caption{The pressure at critical temperature
$T_{cr}$ as a function of compression $\rho/\rho_0$ calculated for
different values of the nonextensivity parameter $q$. The dots
indicate positions of the inflection points for which first
derivative of pressure in compression vanishes. As in \cite{Sousa}
for $q = 1$ the corresponding compression is $\rho/\rho_0 = 1.67$
(and this leads to $\mu = 318$ MeV); it remains the same for $q
> 1$ considered here (but now $\mu = 321$ MeV for $q = 1.01$ and
$\mu = 325$ MeV for $q = 1.02$)  whereas it is shifted to
$\rho/\rho = 1.72$ for $ q< 1$ (resulting in $\mu = 313$ MeV for
$q = 0.99$ and $\mu = 307$ MeV for $q = 0.98$).} \label{FIG:4}
\end{figure}

\begin{figure}[h]
\begin{center}
\includegraphics[width=7.7cm]{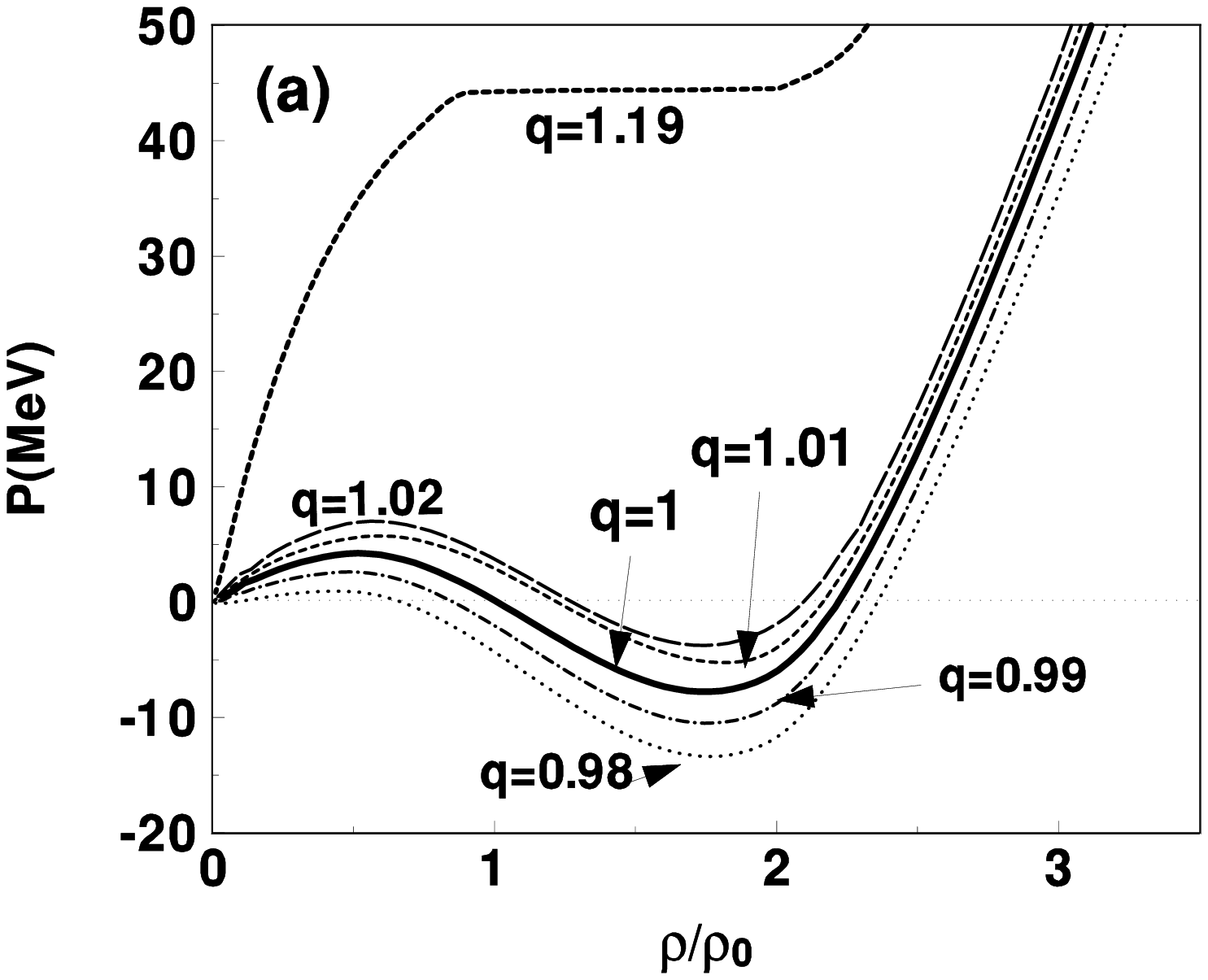}
\includegraphics[width=7.7cm]{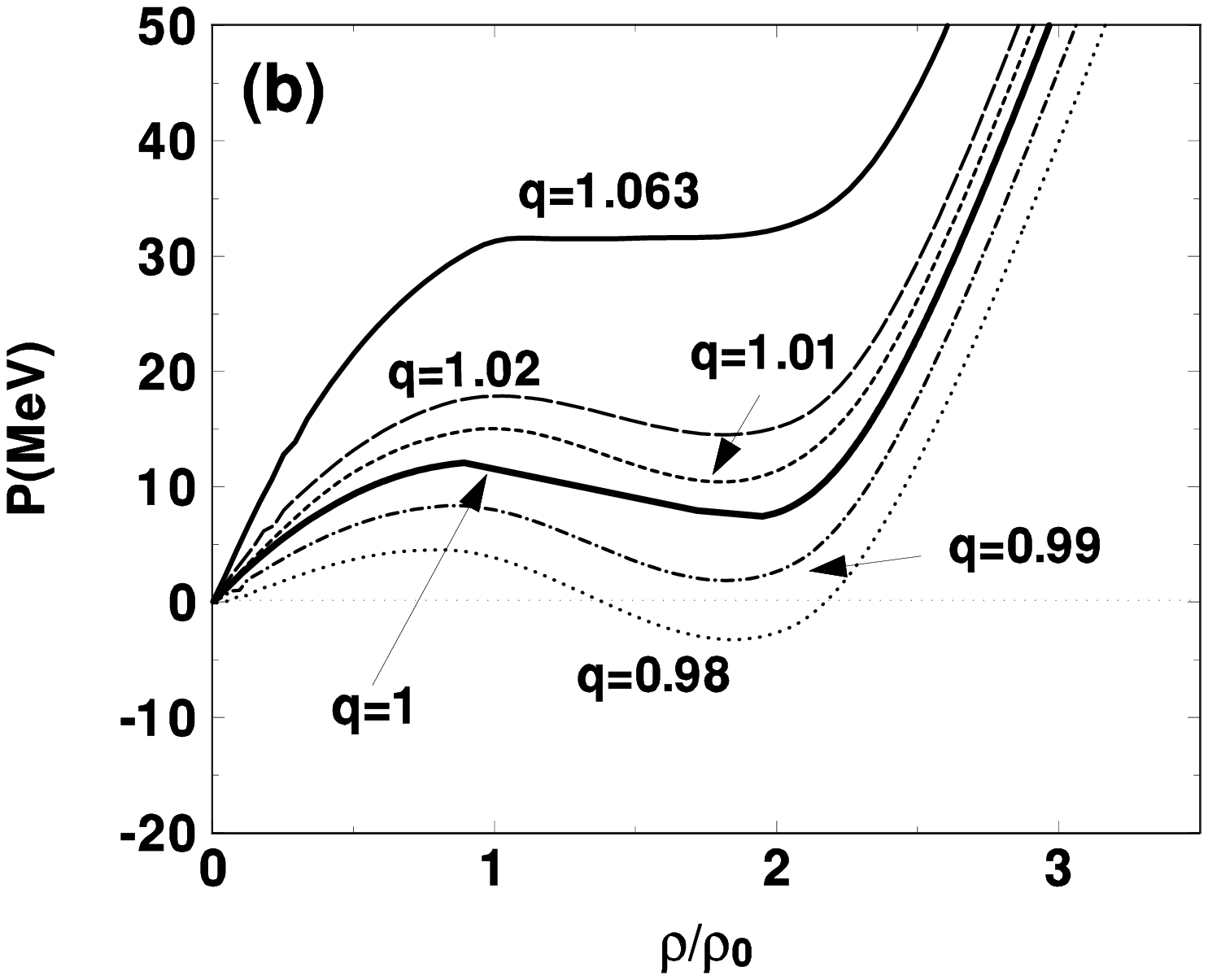}
\end{center}
\vspace{-0.5cm} \caption{The pressure calculated for different
values of the nonextensivity parameter $q$ for temperatures $T =
30$ Mev (a) and $T = 50$ MeV (b) as function of the compression
$\rho/\rho_0$. The curves for $q$ for which the temperature
considered is the critical temperature are also shown, they
correspond to $q=1.19$ for $T=30$ MeV and $q=1.063$ for $T = 50$
MeV.} \label{FIG:5}
\end{figure}

\begin{figure}[h]
\begin{center}
\includegraphics[width=9cm]{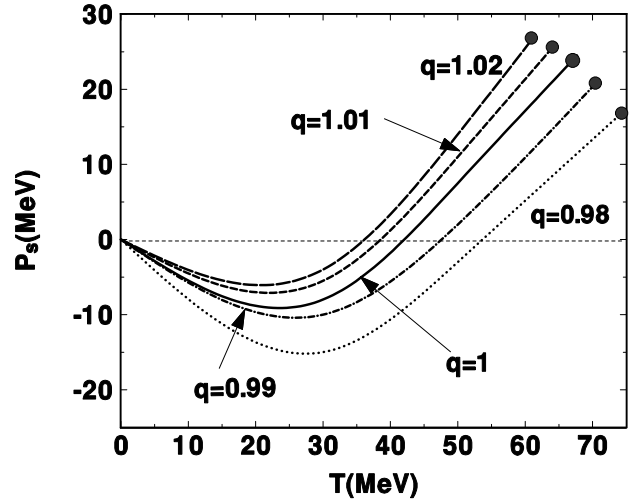}\\
\end{center}
\vspace{-0.5cm} \caption{Pressure $P_s$ corresponding to the local
minimum in Figs. \ref{FIG:4} and \ref{FIG:5} versus temperature
$T$ calculated for some selected values of the nonextensivity
parameter $q$. The curves end at the critical points.}
\label{FIG:6}
\end{figure}

\begin{figure}[t]
\begin{center}
\includegraphics[width=7.7cm]{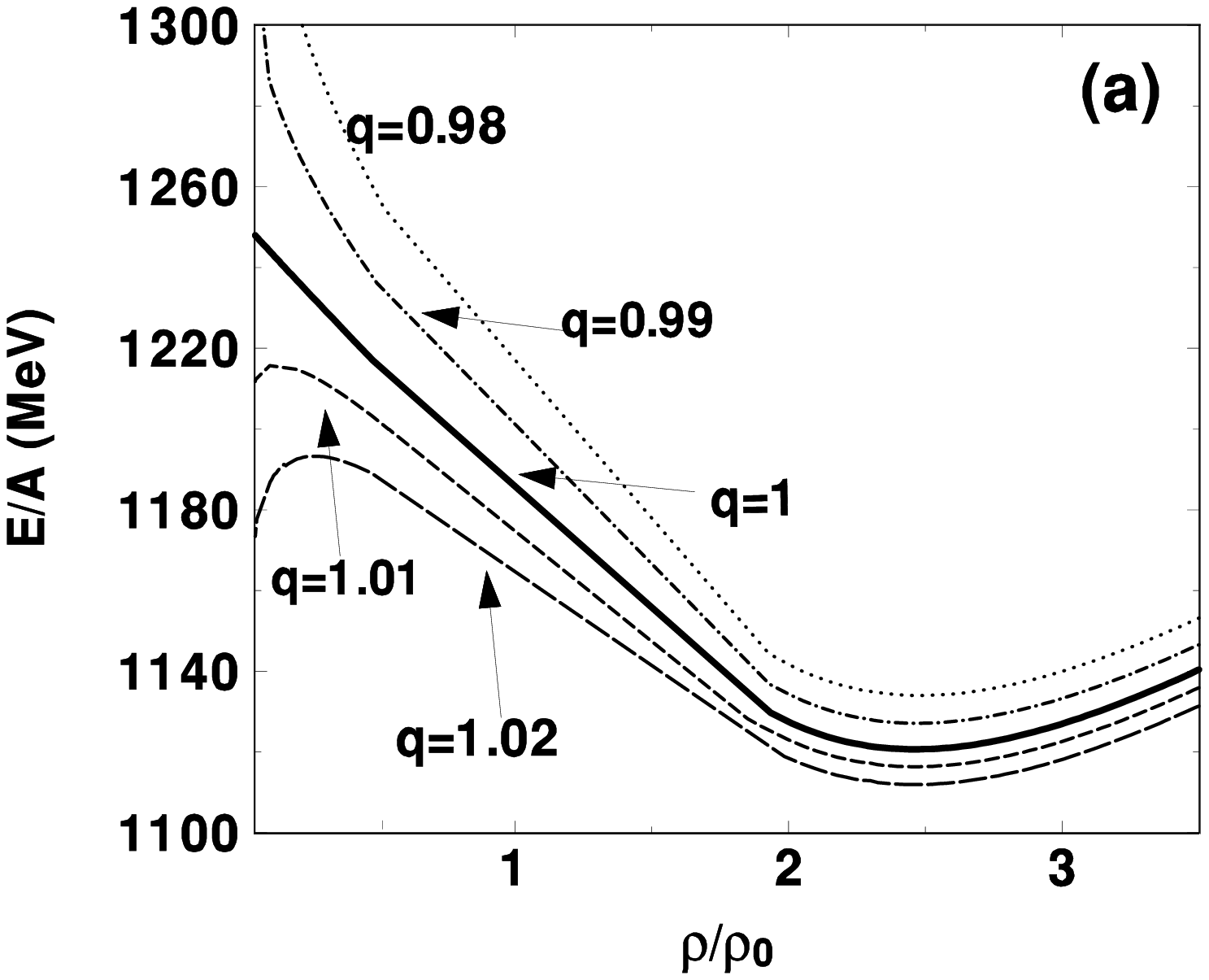}
\includegraphics[width=7.7cm]{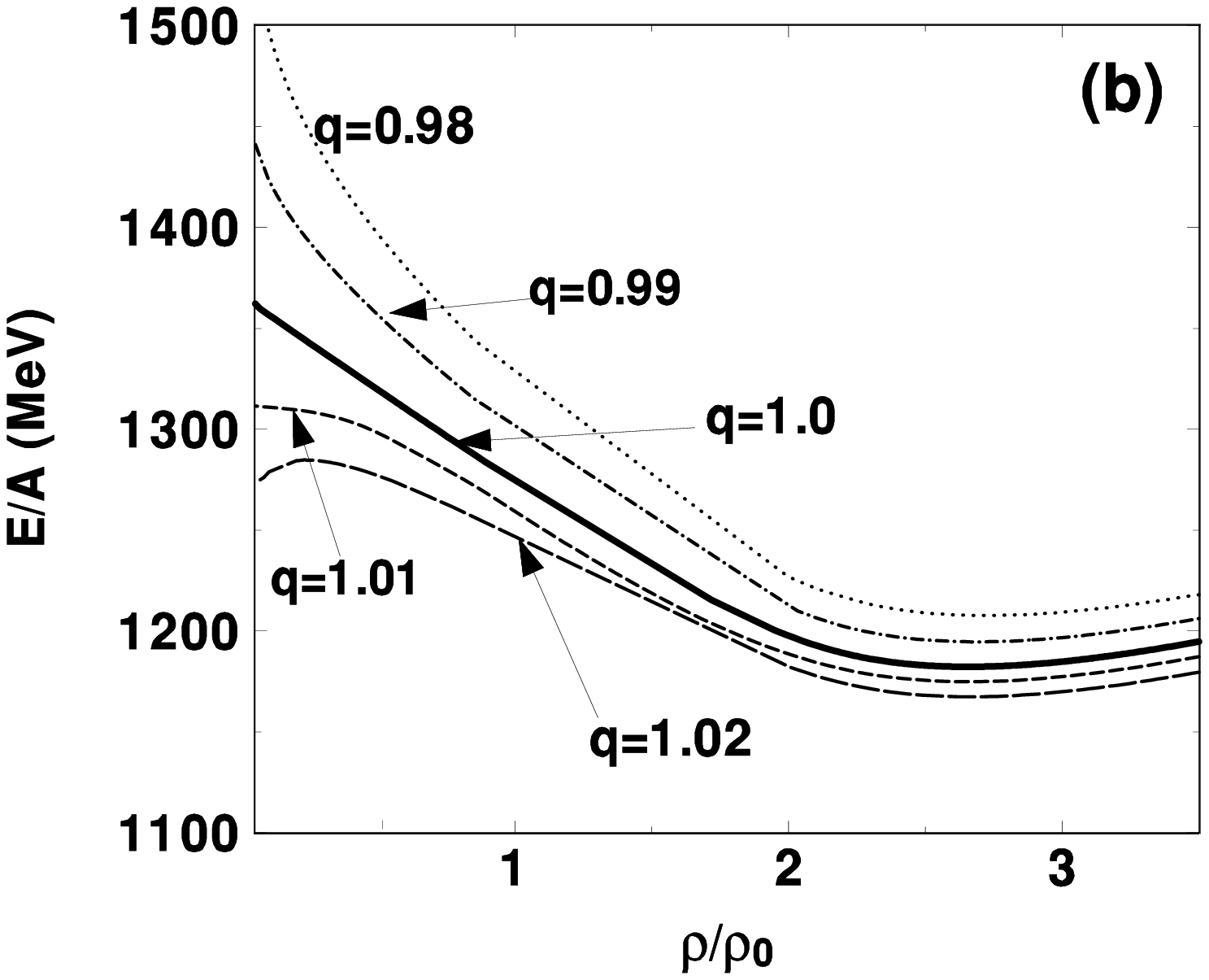}
\end{center}
\vspace{-0.5cm} \caption{The energy per particle at the
temperatures $T = 30$ and $T = 50$ MeV as a function of
compression $\rho/\rho_0$ calculated for different values of the
nonextensivity parameter $q$. } \label{FIG:7}
\end{figure}

The spinodal phase transition occurs, in  general, for finite
densities and for temperatures below a critical temperature
$T_{cr}$ \cite{spinodal,departure}. Above it we do not observe
phase transition of the first order but rather a smooth crossover.
Below it, for some range of densities, we have a region of mixed
phases of hadronic and quark matter \cite{Sousa} (understood here
as phases with very small (current) and very large (constituent)
quark masses). The first observation is that details of the
spinodal phase transition are very sensitive to $q - 1$, much more
than it was observed in the previous case. It is best seen in the
$q$-dependence of $T_{cr}$ shown in Fig. \ref{Fig:3}, it changes
by $\sim 10$ MeV between $q = 0.98$ and $q = 1.02$ used here. The
general observation is that for $q < 1$ the pressure decreases and
energy increases, whereas for $q > 1$ one observes the opposite
tendency, cf. Figs. \ref{FIG:4} - \ref{FIG:7}. Such behavior is a
direct consequence of the nonextensivity and arises from the the
$(q - 1)$ term in the nonextensive entropy functional
(\ref{eq:S-q}).

In Fig. \ref{FIG:4} we show the pressure at the critical
temperatures $T_{cr}$ for different values of $q$ as a function of
the compression $\rho/\rho_0$ (with $\rho_0 = 0.17$ fm$^{-3}$).
Notice that the effect is stronger for $ q <1$ and that,
essentially, the saddle point remains at the same value of
compression. When one moves away from the critical temperature,
one gets the pressure versus compression as presented in Fig.
\ref{FIG:5} for two different temperatures, $T = 30$ and $50$ MeV.
Notice the occurrence of the typical spinodal structure, which is
more pronounced for lower temperatures, $T = 30$ MeV, whereas its
sensitivity to the $q$ parameter gets stronger with increasing
temperature. Shown are also curves for $q$'s for which the
temperature considered coincides with the critical temperature for
this value of $q$. The are, respectively, $q = 1.19$ for $T = 30$
MeV and $q = 1.063$ for $T = 50$ MeV, i.e., the corresponding
values of $q$ decrease with temperature, as expected from Fig.
\ref{Fig:3}. It means that for each temperature (even for very
small one) a $q > 1$ exists for which there is no more mixed phase
and for which spinodal effect vanishes. This seems to be a quite
natural effect in the scenario in which $q > 1$ is attributed to
the fluctuations of the temperature in a system considered as
proposed in \cite{WW,EPJA}. On the contrary, effects like
correlations or limitations of the phase space considered in
\cite{Kodama,fractal} work towards an increase of the $T_{cr}$ and
make therefore the spinodal effect more pronounced.

Fig. \ref{FIG:6} shows the temperature behavior of the pressure
$P_s$ defined as the pressure at the local minimum in the spinodal
curve. The characteristic features are that the temperature at
which $P_s$ starts to be positive and is shifted towards smaller
values with increasing $q$ and that it gets deeper into negative
values with decreasing $q$. The effect of negative pressure can be
best understood invoking the bag model picture of the nucleon
\cite{Bub}. Generally, in the phase of hadron gas we observe a
decrease of the pressure and the critical temperature $T_{cr}$
with the increase of fluctuations given by $q$. That phenomenon
resembles to some extent the behavior of the bag constant for
nucleon in the medium where the bag pressure decreases with the
increase of the chemical potential $\mu$ in order to get a proper
equation of state \cite{bags}. In the nuclear thermodynamical
models this bag constant is modified because the vacuum, in which
hadrons are embedded, is modified by the residual interaction
present in the nuclear medium (acting towards the Wigner
realization of the chiral symmetry in which masses of $\pi$ and
$\sigma$ are degenerated). Here such a density dependence
corrections are introduced by nonextensive effects inside the
nuclear medium. In that way the nuclear vacuum for the
temperatures below the critical temperature and critical
densities, the usual area of the spinodal phase transition, can be
properly described effectively by the nonextensive statistics.

Fig. \ref{FIG:7} shows the energy per particle, $E/A$ (cf. Eq.
({\ref{q_energy}) for different temperatures and for different
values of the nonextensivity parameter $q$. Notice that whereas
for $q < 1$ this energy exceeds the usual one (i.e., for $ q = 1$)
it gets smaller for $q >1$ (this effect is especially seen for
compressions smaller than two). This is an opposite trend to that
observed for the corresponding behavior of the pressure. Finally,
it is worth to observe that the absolute minimum of energy for
given temperature does not depend on the nonextensivity parameter
$q$ and for $T = 30$ MeV it is located at $\rho/\rho_0 = 2.45$. It
turns out that to obtain stable state here, i.e., $P = 0$, one has
to choose $q = 0.97$. In such a way the final droplets of quarks
\cite{BO} in the mixed phase can appear at finite temperatures.

\begin{figure}[t]
\begin{center}
\includegraphics[width=9cm]{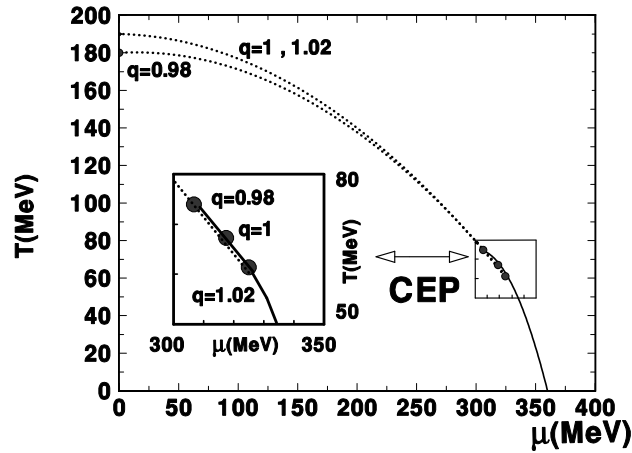}
\end{center}
\vspace{-0.5cm} \caption{Phase diagram in the $q$-NJL model in $T
- \mu$ plane for values of $q$ considered before:
$q=0.98,~1.0~,~1.02$. Solid and dashed lines denote, respectively,
the first order and crossover phase transitions. The results are
presented for three different values of the nonextensivity
parameter $q$ with the vicinity of the ($q$-dependent) critical
end points (CEP) enlarged in the inlet. The crossover phase
transition for $q = 0.98$ and for $\mu \rightarrow 0$ takes place
for smaller temperature $T$.} \label{Fig:8}
\end{figure}

Let us close with few remarks. Nonextensive dynamics enter the NJL
calculations through the quark (antiquark) number distribution
functions $n_{qi}$ ($\bar{n}_{qi}$). These functions are connected
with the respective quark (antiquarks) spectral functions in NJL
model. However, deviations from the exponential shape of
$q$-exponents, as defined in Eqs. (\ref{q>1}) and (\ref{q<1}), are
negligible for values of $q$ close to unity (in our case
$0.98<q<1.02$). It is also important to notice that Eqs.
(\ref{q>1}) and (\ref{q<1}) are symmetric for $q$ and $1-q$. The
differences between $q<1$ and $q>1$ cases observed in our results
are then due to our way of defining the energy (\ref{q_energy})
and entropy (\ref{q entropy}), which, following \cite{Drago,AL},
we do by using $n^q{_{qi}}$ and $\bar{n}^q{_{qi}}$ instead of
$n_{qi}$ and $\bar{n}_{qi}$ \footnote{It is worth to notice that
in \cite{Pereira}, which considers only the $q>1$ case and uses
number distributions without powers of $q$, the significant
effects were obtained only for much bigger values of the
nonextensive parameter $q=1.2$.}. Because now for $q<1$
distributions $n^q{_{qi}}$ and $\bar{n}^q{_{qi}}$ are closer to
unity than $n{_{qi}}$ and $\bar{n}{_{qi}}$, therefore the absolute
values of quark condensates (as given by Eq. (\ref{q_gap1})) begin
to decrease for $q=0.98$ at lower temperature  in comparison with
the $q=1$ case, see Fig. \ref{Fig:1}. The corresponding energy,
see Fig. \ref{FIG:7}, is bigger, which means that $q < 1$
introduces some residual attractive correlations which rise the
energy and lead to hadronization occurring at lower temperature
(see Figs. \ref{Fig:1} and \ref{Fig:2}). On the other hand, $ q >
1$ introduces fluctuations which decrease the effective
occupations ($n^q{_{qi}}$ and $\bar{n}^q{_{qi}}$) and the energy,
and smears out the chiral phase transition, see Fig. \ref{Fig:2}.
In Fig. \ref{Fig:8} we present our phase diagram in the $\mu-T$
plane for different nonextensivity parameters considered here with
positions of the corresponding critical end points (CEP) for
different values of $q$ clearly indicated.  The overlap of curves
observed in Fig. \ref{Fig:8} (inlet) indicates how critical end
point is smeared to a kind of critical area. This is because
fireballs created in different events can have different values of
$q$ (representing, as mentioned before, action of all factors
responsible for the departure of our system from the usual BG
approach - not specified here in detail but, in general, resulting
in specific correlations of quarks or fluctuations of temperature
mentioned before). Therefore when analyzing experimental data one
most probably will encounter such a critical area instead of well
defined CEP.

\section{\label{sec:IV}Summary and conclusions}

To summarize: we have investigated the sensitivity of the mean
field theory of the NJL type presented in \cite{Sousa} to the
departure from the conditions required by the application of the
BG approach. To this end we have used the Tsallis version of
nonextensive statistical mechanics \cite{T} with a new parameter
$q$, the phenomenological nonextensivity parameter, such that
$q-1$ quantifies departure from the BG situation (which is
recovered for $q \to 1$). As result we have obtained a
nonextensive, $q$-NJL model with $q$ being a new parameter
summarizing action of a number of yet undisclosed factors, which
should be fitted to the data. Our investigation was prompted by
recent investigations of similar effects in quantum hadrodynamics
\cite{Pereira}. In fact, we have used the same nonextensive
version of the standard FD distributions (discussed in
\cite{TPM}), but this time applied to quarks and antiquarks and
for both $q > 1$ and $ q < 1$ cases. On the other hand, when
calculating energies, quark condensates and densities we followed
prescription advocated in \cite{Drago,AL}.

We have investigated two possible scenarios corresponding to $q
> 1$ and $q < 1$, respectively, which, as mentioned, correspond to
different physical interpretations of the nonextensivity
parameter. For $ q <1$ (usually connected with some specific
correlations \cite{Kodama} or with fractal character of the phase
space \cite{fractal}) we observe decreasing of pressure, which
reaches negative values for a broad ($q$-dependent) range of
temperatures and increasing of the critical temperature
\footnote{It acts therefore in the same way as including of the
Polyakov loop into the NJL model \cite{PNJL}.}.  The $ q > 1$ case
(usually connected with some specific nonstatistical fluctuations
existing in the system \cite{WW,EPJA} \footnote{Actually, in
\cite{qH} it was shown that if one could find a dynamical source
of such fluctuations and introduce it to the model then the new
$q$ would diminish considerably, eventually becoming unity.}) we
observe a decreasing of the critical temperature, $T_{crit}$ (cf.
Fig. \ref{Fig:3}), and therefore in the limit of large $q$ we do
not have a mixed phase but rather a quark gas in the deconfined
phase above the critical line (on the contrary, the compression at
critical temperature does not depend on $q$ (cf. Fig.
\ref{FIG:4}). As in \cite{Pereira} the resulting equation of state
is stiffer (in the sense that for a given density we get bigger
pressure with increasing $q$). As expected, the observed effects
depend on the temperature and tend to vanish when temperature
approaches zero. Finally, as shown in Fig. \ref{Fig:8}, the
nonequilibrium statistics dilutes the border between the crossover
and the first order transition.

We would like to end by stressing that our results could be of
interest for investigations aimed at finding the critical point in
high energy heavy ion collisions \cite{departure} or when studying
particularities of the equation of state (EoS) of compact stars
\cite{NSTARS}. The fact that they do depend on the parameter $q$
used means that the exact position of such a point or the type of
the or the shape of EoS could be quite different from that naively
expected.

\section*{Acknowledgment}

Partial support (GW) of the Ministry of Science and Higher
Education under contract 1P03B02230 is gratefully acknowledged.


\section*{References}
\begin{thebibliography}{99}

\bibitem{departure} Randrup J 2009 {\it Phys. Rev.} C {\bf 79}
                    054911;
                    Palhares L F, Fraga E S, and Kodama T {\it
                    Finite-size effects, pseudocritical quantities
                    and signatures of the chiral critical endpoint
                    of QCD}, arXov:0904.4830[nucl-th].
                    See also Skokov V V,and Voskresensky D N,
                    {\it Hydrodynamical description of a hadron-quark
                    first-order phase transition},
                    arXiv:0811.3868[nucl-th] and 2009
                    {\it Nucl. Phys.} A {\bf 828} 401 and references therein.
\bibitem{T} Tsallis C, J. 1988 {\it Stat. Phys.} {\bf 52} 479;
            Salinas S R A, and Tsallis C (eds.), 1999
            {\it Special Issue on Nonextensive Statistical Mechanics
            and Thermodynamic}, {\it Braz. J. Phys.} {\bf 29};
            M. Gell-Mann M, and Tsallis C (Eds.) 2004
            {\it Nonextensive Entropy Interdisciplinary
            Applications} (Oxford University Press, New York);
            Tsallis C 2009 {\it Eur. Phys. J.} A {\bf 40} 257

\bibitem{applications}  Gell-Mann M and Tsallis C 2004 (Eds.)
                        {\it Nonextensive Entropy - interdisciplinary
                        applications} (Oxford University Press);
                        Boon J P, and Tsallis C (Eds.) 2005 {\it Nonextensive
                        Satistical Mechanics: New Trends, New Perspectives},
                        {\it Europhysics News} {\bf 36} (special issue).
                        An updated bibliography on Tsallis's nonextensive statistics can
                        be found at http://tsallis.cat.cbpf.br/biblio.htm.

\bibitem{EPJA} Wilk G, and W\l odarczyk Z 2009 {\it Eur. Phys. J.} A {\bf
               40} 299

\bibitem{AL} Alberico W M, and Lavagno A 2009 {\it Eur. Phys. J.} A
            {\bf 40} 313

\bibitem{qHydro} Osada T, and Wilk G 2008 {\it Phys. Rev.} C {\bf 77} 044903
                 and 2009 {\it Centr. Eur. J. Phys.} {\bf 7} 432

\bibitem{qH} Biyajima M, Mizoguchi T, Nakajima N, Suzuki N, and Wilk G
             2006 {\it Eur. Phys. J.}  C {\bf 48} 597.

\bibitem{qBiro} Bir\'{o} T S, and Purcsel G 2009 {\it Centr. Eur. J. Phys.} {\bf 7} 395;
                Bir\'{o} T S, Purcsel G, and \"Urmo\"sy K 2009
                {\it  Eur. Phys. J.} A  {\bf 40}   325

\bibitem{Drago} Drago A, Lavagno A, and Quarati P 2004 {\it Physica} A {\bf 344} 472

\bibitem{stationary} Bir\'{o} T S, and Purcsel G 2005 {\it Phys. Rev. Lett.} {\bf 95}
                     162302 and 2008 {\it Phys. Lett.}  A {\bf 372},
                     1174; Bir\'{o} T S 2008 {\it Europhys. Lett.}
                     {\bf 84} 56003

\bibitem{Kaniadakis} Kaniadakis G 2009 {\it Eur. Phys. J.} A  {\bf 40} 275

\bibitem{MaxEnt_K} Kaniadakis G 2009 {\it Eur. Phys. J.} B {\bf 70} 3

\bibitem{todos}  Hasegawa H 2009 {\it Physica} A {\bf 388} 2781 and 2009 {\it Phys.
                 Rev.} E {\bf 80} 011126;
                 Reis M S, Amaral V S, Sarthour R S, and Oliveira I S 2006
                 {\it Phys. Rev.} B {\bf 73} 092401;
                 Douglas P, Bergamini S, and Renzoni F, 2006 {\it Phys. Rev. Lett.}
                 {\bf 96} 110601;
                 Silva R, Fran\c ca G S, Vilar C, and Alcaniz J S,
                 2006 {\it Phys. Rev.} E {\bf 73} 026102;
                 Jiulin Du 2004 {\it Europhys. Lett.} {\bf 67} 893;
                 Silva R, and Alcaniz J S 2004 {\it Physica} A {\bf 341}
                 208 and 2003 {\it Phys. Lett.} A {\bf 313} 393

\bibitem{WW} Wilk G, and W\l odarczyk Z 2000 {\it Phys. Rev. Lett.} {\bf 84} 1770
             and 2001 {\it Chaos, Solitons and Fractals} {\bf 13/3} 581

\bibitem{BJ} Bir\'{o} T S, and Jakov\'ac A 2005 {\it Phys. Rev.
             Lett.} {\bf 94} 132302

\bibitem{Kodama} Kodama T, Elze H-T, Aguiar C E, and Koide T 2005
                 {\it Europhys. Lett.} {\bf 70} 439;
                 Kodama T, and Koide T 2009 {\it Eur. Phys. J.}
                 A {\bf 40} 289

\bibitem{fractal} Garc\'ia-Morales V, and Pellicer J 2006 {\it Physica} A
                  {\bf 361} 161

\bibitem{TPM} Teweldeberhan A M, Plastino A R, and Miller H C 2005 {\it Phys.
              Lett.} A {\bf 343} 71

\bibitem{TPM1} Teweldeberhan A M, Miller H G, and Tegen R 2003 {\it Int. J. Mod. Phys.}
               E {\bf 12} 395.

\bibitem{DBG} B\"u\"yukkili\c c F, and Demirhan D 1993 {\it Phys Lett.} A {\bf 181}
              24;
              B\"uy\"ukkili\c c F, Demirhan D, and G\"ule\c c A 1993 {\it Phys.
              Lett.} A {\bf 197} 209

\bibitem{SW} Walecka J D 1974 {\it Ann. Phys.} {\bf 83} 491;
             S. A. Chin S A, and Walecka J D, 1074 {\it Phys. Lett.} B
             {\bf 52} 1074;
             Serot B D, and Walecka J D 1986 {\it Advances in Nuclear
             Physics} {\bf 16} (Plenum Press, New York)

\bibitem{Pereira} Pereira F I M, Silva B, and Alcaniz J S,
                  2007 {\it Phys. Rev.} C {\bf 76} 015201

\bibitem{Sousa} Costa P, Ruivo M C, and de Sousa A 2008 {\it Phys. Rev.} D {\bf 77},
                096001

\bibitem{Bernard1} Bernard V, Jaffe R L, and Meissner U-G 1988 {\it
                   Nucl. Phys.} B {\bf 308}, 753

\bibitem{Bernard2} Bernard V, Osipov A A, and Meissner U-G 1992 {\it
                   Phys. Lett.} B {\bf 285}, 119

\bibitem{NJL} Nambu Y, and Jona-Lasinio G 1961 {\it Phys. Rev.} {\bf 122} 345
              and 1961 {\it Phys. Rev.} {\bf 124} 246; see also:
              Rehberg P, Klevansky S P, and H\"ufner J 1996 {\it Phys. Rev.} C
              {\bf 53} 410

\bibitem{HK} Hatsuda T, and Kunihiro T 1994 {\it Phys. Rep.} {\bf  247} 221
             and references therein; cf. also Bernard V, Meissner U -G,
             and Zahed I 1987 {\it Phys. Rev.} D {\bf 36} 819

\bibitem{Kle} Klevansky S P 1992 {\it Rev. Mod. Phys.} {\bf 64} 649

\bibitem{Bub} Buballa M 2005 {\it Phys. Rep.} {\bf 407} 205

\bibitem{Costa} Costa P, Ruivo M C, de Sousa A, and Kalinovsky Yu L
                2004 {\it Phys.Rev.} C {\bf 70} 025204

\bibitem{Gross} Gross D H E 2004 {\it Physica} A {\bf 340} 76,
                2006 {\it Lecture Notes in Physics} (Springer) {\bf 602} 23
                and 2006 {\it Physica} A {\bf 365} 138;
                see also Gross D H E 2001 {\it Microcanonical thermodynamics,
                Phase transitions in "Small Systems"}, WS Lecture
                Notes in Physics - Vol. 66, World Sci.

\bibitem{Tq} Tsallis C, Rapisarda A, Latora V, and Baldovin F
            (2002) {\it Lecture Notes in Physics} (Springer) {\bf 602} 140

\bibitem{NegC} Rapisarda A, and Latora V {\it Negative specific heat
                in out-of-equilibrium nonextensive systems},
                arXiv:nucl-th/0202075; Gross D H E 2006 {\it Physica} A {\bf 365}
                138 (and references therein)

\bibitem{tsallis05} Tsallis C, Gell-Mann M, and Sato Y 2005 {\it Proc. Nat. Acad.
                    Sci.} {\bf 120} 15377

\bibitem{Abe} Abe S 2006 {\it Physica} A {\bf 368} 430

\bibitem{spinodal} Randrup J 2004 {\it Phys. Rev. Lett.} {\bf 92} 122301

\bibitem{bags} Leonidov A, Redlich K, Satz H, Suhonen E, and Weber G
               1994 {\it Phys. Rev.} D {\bf 50} 4657;
               Patra B K, and Singh C P 1996 {\it Phys. Rev.} D {\bf 54}
               3551

\bibitem{BO} Buballa M, and Oertel M (1998) {\it Nucl. Phys.} A {bf 642} 39
             and 1999 {\it Phys. Lett.} {\bf B457} 251


\bibitem{PNJL} Costa P, de Sousa C A, Ruivo M R, and Hansen H 2009
               {\it Europen Phys. Lett.} {\bf 86} 31001

\bibitem{NSTARS} Kl\"ahn T, Blaschke D, Typel S, van Dalen E N E,
                 Faessler A, Fuchs C, Gaitanos T, Grigorian H,
                 Ho A, Kolomeitsev E E, Miller M C, R\"opke G,
                 Tr\"umper J, Voskresensky D N, Weber F, and
                 Wolter H H 2006 {\it Phys. Rev.} C {\bf 74} 035802


\end{thebibliography}
\end{document}